\documentclass[twocolumn,showpacs,amsmath,amssymb,prb,superscriptaddress,reprint]{revtex4-1}
\usepackage{amsmath}
\usepackage[english]{babel}
\usepackage{graphicx}% Include figure files
\usepackage{bm}% bold math

\begin{document}

%\title{Interplay of charge and vorticity quantization in superconducting single electron devices}
\title{Interplay of charge and vorticity quantization in superconducting Coulomb blockaded island}

\author{ I.M.\ Khaymovich}
\affiliation{Low Temperature Laboratory, Department of Applied Physics, Aalto University, P. O. Box 13500, FI-00076 AALTO, Finland}
\affiliation{Institute for Physics of Microstructures, Russian
Academy of Sciences, 603950 Nizhni Novgorod, GSP-105, Russia }

\author{ V.F.\ Maisi}
\affiliation{Solid State Physics Laboratory, ETH Z{\"u}rich, 8093 Z{\"u}rich, Switzerland}
\affiliation{Centre for Metrology and Accreditation (MIKES), P.O. Box 9, 02151 Espoo, Finland}
\affiliation{Low Temperature Laboratory, O.V. Lounasmaa Laboratory, Aalto University, P. O. Box 13500, FI-00076 AALTO, Finland}

\author{J.P.\ Pekola}
\affiliation{Low Temperature Laboratory, O.V. Lounasmaa Laboratory, Aalto University, P. O. Box 13500, FI-00076 AALTO, Finland}

\author{A.S.\ Mel'nikov}
\affiliation{Institute for Physics of Microstructures, Russian
Academy of Sciences, 603950 Nizhni Novgorod, GSP-105, Russia }
\affiliation{Lobachevsky State University of Nizhni Novgorod, 23 Prospekt Gagarina, 603950, Nizhni Novgorod, Russia}

\date{\today}

\begin{abstract}
%overlooked concept
The angular momentum or vorticity of Cooper pairs is shown to affect strongly the charge transfer through a
small superconducting (S) island of a single electron transistor.
This interplay of charge and rotational degrees of freedom
in a mesoscopic superconductor occurs through the effect of vorticity on the quantum mechanical spectrum
of electron-hole excitations. The subgap quasiparticle levels in vortices can host an additional electron, thus, suppressing
the so-called parity effect in the S island. We propose to measure this interaction between the quantized
vorticity and electric charge via the charge pumping effect caused by alternating vortex entry and exit controlled by a periodic
magnetic field.
%This magnetoelectric phenomenon is shown to
% provide an intriguing possibility to confirm the discrete nature of subgap energy levels
%and detect the value of the minigap in the vortex core spectrum.
%
\end{abstract}

\pacs{85.35.Gv, 73.23.Hk, 74.25.Ha}
%85.35.Gv	Single electron devices
%73.23.Hk	Coulomb blockade; single-electron tunneling
%74.25.Ha	Magnetic properties including vortex structures and related phenomena
%85.80.Jm	Magnetoelectric devices
%75.85.+t	Magnetoelectric effects, multiferroics
%74.78.Na	Mesoscopic and nanoscale systems (in 74.78.-w	Superconducting films and low-dimensional structures)
%74.45.+c	Proximity effects; Andreev reflection; SN and SNS junctions
\maketitle

There are two fundamental quantization phenomena which are manifested in different aspects of physics of superconducting (S) metals:
(i) quantization of charge of superconducting carriers or Cooper pairs in units of double electron charge
 $2e$  and (ii) quantization of trapped magnetic flux in units of the flux quantum
$\Phi_0 =h c/2e$ %\pi\hbar c/e$
which originates from the quantization of the angular momentum of Cooper pairs or vorticity.
Basic superconducting theory shows that these two quantization rules cannot be considered independently.\cite{Tinkham} %\cite{Charge-flux_Cleary,Charge-flux_Kulik,Charge-flux_MelPRL}.
%One can naively expect that the corresponding magnetoelectric phenomena could manifest themselves in a
Certainly it would be exciting to clarify if this interplay of charge and vorticity quantization can reveal itself in possible magnetoelectric phenomena inherent, e.g., to a superconducting state containing topological
defects with quantized vorticity, namely, Abrikosov vortices.
%Indeed,
%taking, e.g., a rotating superconducting condensate in a single Abrikosov vortex, we immediately destroy the Cooper pairs in the close
%vicinity of the core of the vortex. As a result, one can naively expect that the charge inside the core is quantized in units of $e$.
Unfortunately for the bulk samples it is extremely difficult to observe these effects %this interplay of charge and vorticity quantization
experimentally.
The reason is that for typical metals the large value of the Fermi energy $E_F$ almost completely suppresses all the electrostatic charge
phenomena caused by vortices: the vortex core charge is small due to the very small ratio
$\Delta/E_F\sim 10^{-5}-10^{-2}$.\cite{Vortex_charge_GeshkenbeinPRL,Vortex_charge_KhomskiiPRL} Here $\Delta$ is the superconducting order parameter.

%%%%%%%%%%%%%%%%%%%%%%%%%%%%%%%%%%%%%%%%%%%%%%%%%%%%%%%%%%%%%%%%%%%%
\begin{figure}[t]
\includegraphics[width=0.33\textwidth]{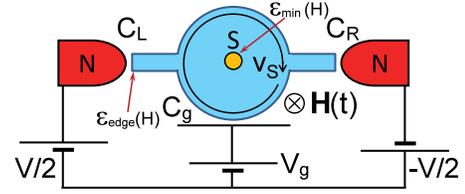}
\caption{(color online) Setup of the NISIN
SET with a bias voltage $\pm V/2$ applied to the normal metal electrodes tunnel coupled to the central S disc with capacitances $C_L$ and $C_R$.
%The gate voltage $V_g=e n_g/C_g$ is applied to the capacitively coupled gate electrode with the capacitance $C_g$.
Magnetic field is
applied perpendicular to the disc plane.
}
 \label{Fig:setup}
\end{figure}
%%%%%%%%%%%%%%%%%%%%%%%%%%%%%%%%%%%%%%%%%%%%%%%%%%%%%%%%%%%%%%%%%%%%

In the present Letter we dare to make a suggestion overcoming the above difficulties based on
the powerful methods provided by superconducting single electron devices (see, e.g., Refs. \onlinecite{Averin_Likharev86+91,Pekola_RMP} and references therein).
The key point of this idea %can be formulated quite briefly.
is that by creating or removing a single vortex in a small superconducting island, either the odd or even electron number will be favored. Thus the vorticity and the charge of the island will be coupled.
%To address electrons one by one we propose to use a single electron transistor (SET), i.e., a small metallic island with
%total electric capacitance $C=C_L+C_R+C_g$ (see Fig.~\ref{Fig:setup}) in the regime of Coulomb blockade. The island is coupled to the leads by tunnel contacts.
To address electrons one by one we propose to use a single electron transistor (SET), i.e., a small metallic Coulomb-blockaded island with
total electric capacitance $C=C_L+C_R+C_g$ (see Fig.~\ref{Fig:setup}) coupled to the leads by tunnel contacts.
Large Coulomb energy $E_C=e^2/(2C)$ of the island compared to the temperature $T$ prevents an additional electron to tunnel in and allows one to manipulate the charge state $n$ of the island in a controllable way by varying the electrical potential of the gate electrode $V_g=e n_g/C_g$ (see Fig.~\ref{Fig:setup}), where $C_g$ is the gate electrode capacitance.
For an S island this physical picture becomes more complicated due to the %so~-~called
electron number parity effect.\cite{Tuominen_Tinkham,Averin_Nazarov,NISIN_parity_Devoret,NISIN_parity_Martinis}
This %parity
effect consists in the $2e$~-~periodic dependence of the observables on an applied gate voltage and provides, thus, a
direct confirmation of the Cooper pair charge quantization.
At finite temperatures the parity phenomena are
controlled by the free energy difference $\delta F = \Delta - T \ln N_{eff}$ between the states with
odd and even number of excess electrons in the granule. Here  $N_{eff}$ equals the effective number of available states for an additional particle.
It is clear that the parity effect is observable only for a positive value of this free energy barrier, i.e., at low enough temperatures:
 $T<T^{*}=\Delta/\ln N_{eff}$.
 Applying an external magnetic field one can suppress partially the S gap and, thus, suppress the parity phenomenon.
 One can observe this suppression either by measuring the change of periodicity of the SET characteristics vs the gate voltage
in an applied magnetic field $H$,\cite{Parity_effect_in_H}
or by using a varying magnetic field for controlled electron transfer through the SET at a fixed gate potential.
At low temperatures the guaranteed suppression of the parity effect can be achieved by introducing a vortex line in the granule which provides a natural trap for an entering electron.

\begin{figure}[t]
\includegraphics[width=0.35\textwidth]{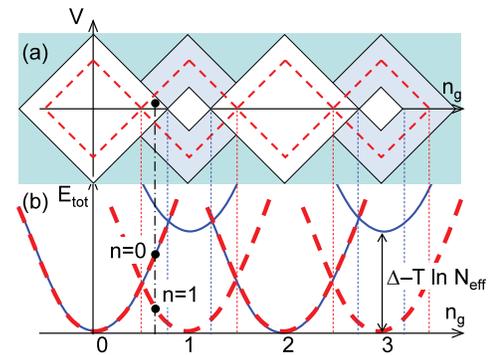}
\caption{(color online)
(a) Stability diagram of the S island in the plane $n_g - V$.
Red dashed lines correspond to the normal state stability diagram.
The zero current plateaus in the regime of the parity effect are shown by the white diamonds.
(b) Ground-state energy of the S island vs gate voltage with (blue solid lines) and without (red dashed lines) parity effect.
%The black dots on all diagrams mark the operational point for the charge state modulation.
}
 \label{Fig:energy_diag}
\end{figure}

Applying an \emph{oscillating} magnetic field, i.e., changing periodically the island vorticity, we can
induce the even-odd transitions in the number $n$ of electrons trapped on the island.
Choosing the gate voltage as shown in Fig.~\ref{Fig:energy_diag} by a black dot, one can switch between the states $n=0$ and $n=1$.
Without the vortex, the even electron number, $n=0$, shown by the large white diamond is favored.
With the vortex, the odd electron number becomes preferable as shown by the dashed red diamonds.
Applying a constant bias voltage to the SET one can convert this modulation of the charge state into unidirectional charge pumping.
The above picture of the vortex controlled parity effect can change at low temperatures less than
the minigap $\omega_0<\Delta$ in the spectrum of quasiparticles trapped in the vortex core.
This new energy scale $\omega_0$
arises from the quantization of the spectrum of single particle excitations confined within the
core by the Cooper pair potential.\cite{CdGM}
Though this minigap in most superconductors is small compared to the bulk gap it can still restore the parity effect.
%Since the publication of the seminal paper \cite{CdGM} by Caroli, de Gennes and Matricon in 1964,
%the discrete nature of these quasiparticle levels in superconducting  vortices remains one of the foci of the physics of vortex matter.
%Recently this phenomenon has attracted considerable interest in the context of zero bias anomaly in scanning tunneling microscopy and
%spectroscopy (STM/STS) experiments in various compounds \cite{oldSTM,MgB2}, in puzzling observations of the electron -- hole quantum mechanics
%in the mixed state of high-temperature superconducting
%cuprates \cite{HTSC}, and in search of the Majorana fermionic states \cite{Majorana}. Despite of its long history
%we have to conclude regrettably that the prediction of discrete levels inside the vortex core has not been confirmed reliably
%by the experimental data. %even for clean samples and ultralow temperatures.
% In particular, the existing STM data on possible observation of the minigap
%$\omega_0$ in the vortex spectra are very controversial probably because of %the specific
%surface effects \cite{Kopn_Khaym_Meln}.
%Thus, new experimental setups are urgently needed to allow for resolving the spectrum discreteness
%and to provide experimental tests of many fascinating ideas presented.
%Interestingly, the above interplay of charge and vorticity quantization which we expect to be observable in SET devices
%can provide a new tool for the detection of the minigap.
The free energy value paid for the even-odd transition in the electron number in the vortex state
can be estimated as $\delta F_v = \omega_0 - T \ln N_{v}$, where $N_{v}\sim k_F L$, $L$ is the length of the vortex line and $k_F$  is the Fermi momentum. The condition $\delta F_v = 0$ gives us the temperature $T^*_v =\omega_0/\ln N_{v}$ separating the regimes
 with $e$ and $2e$ charge periods in the vortex state and, thus, at temperatures below $T^*_v$  the parity effect can be restored.
It is quite useful to emphasize here a simple analogy with the parity effect in the
Josephson junction\cite{Sharov+Zaikin} where $\omega_0$ should be replaced by the minigap that depends on the phase difference between the SC leads.

%%%%%%%%%%%%%%%%%%%%%%%%%%%%%%%%%%%%%%%%%%%%%%%%%%%%%%%%%%%%%%%%%%%%
%%%%%%%%%%%%%%%%%%%%%%%%%%%%%%%%%%%%%%%%%%%%%%%%%%%%%%%%%%%%%%%%%%%%

%  The number of available states in a vortex also differs from the case of homogeneous granule. This quantity for a vortex trap
% can be estimated as a number of modes in the one~-~dimensional quantum wire: Observing
%this crossover temperature by measuring either the periodicity of the current through the SET vs gate voltage or the vortex pumping effect would %confirm the existence of quantized CdGM levels.

We now proceed with the study of an exemplary SET setup (see Fig.~\ref{Fig:setup}) which allows us to illustrate the above charge-vortex interplay. Hereafter we focus on the single electron transport between the normal metal leads
 and do not consider possible magnetic pumping based on the use of
Cooper pair sluices in Josephson systems with S electrodes.\cite{Gasparinetti_H-pump}
The size of the Coulomb blockaded S island is assumed to be of the order of several coherence lengths $\xi$ so that applying an external magnetic field we can introduce at least one vortex in this island.
The electronic transport through this device can be described by a standard rate equation accounting for parity effects.\cite{NISIN_turnstile_overheating} For the sake of simplicity we restrict to a two-level approximation assuming low temperature regime $T\ll E_C$ and taking the gate voltage
interval $0<n_g<1$. The equation for the $n=1$ charge state probability $p_1$ reads:
\begin{equation}\label{rate_eq}
\frac{dp_1}{dt}=\Gamma_{0\to1}p_0-\Gamma_{1\to 0}p_1 \ ,\quad p_0=1-p_1 \ ,
\end{equation}
where $\Gamma_{0\to 1}$ and $\Gamma_{1\to 0}$ are the rates for the electron tunneling into and out of the island, respectively.
These rates are, of course, determined by the sum of contributions coming from the transport through the contacts with
left and right electrodes:
%$\Gamma_{n\to m}=\Gamma_{n\to m}^{L}[U_L] +\Gamma_{n\to m}^{R}[U_R] $, where $\Gamma_{0\to 1}^{j}[U_j] = \Gamma_j^{e}[U_j]$, $\Gamma_{1\to 0}^{j}[U_j] = \Gamma_j^{o}[-U_j]$ and
$\Gamma_{0\to 1}=\Gamma_{L}^e[U_L] +\Gamma_{R}^e[U_R] $ and $\Gamma_{1\to 0}=\Gamma_{L}^o[-U_L] +\Gamma_{R}^o[-U_R] $, where
\begin{equation}\label{Gamma_pm}
\Gamma_{j}^{k}[U]=\frac{1}{e^2 R_{Tj}}\int\limits_{-\infty}^{\infty} \nu_j(\varepsilon)f_N(\varepsilon-U) [1-f_S^{k}(\varepsilon)]d\varepsilon \
\end{equation}
is an increasing function of $U$.
%\begin{subequations}\label{Gamma_pm}
%\begin{align}
%\Gamma_{0\to 1}^{j}[U_j] &=\int\limits_{-\infty}^{\infty}\frac{\nu_j(\varepsilon)}{e^2 R_{Tj}} f_N(\varepsilon-U_j) [1-f_S^{e}(\varepsilon)]d\varepsilon \ , \\
%\Gamma_{1\to 0}^{j}[U_j] &=\int\limits_{-\infty}^{\infty}\frac{\nu_j(\varepsilon)}{e^2 R_{Tj}} [1-f_N(\varepsilon-U_j)] f_S^{o}(\varepsilon)d\varepsilon \ ,
%\end{align}
%\end{subequations}
Here $R_{Tj}$ is the resistance of the $j$th tunneling junction, $\nu_j(\varepsilon)$ is the local density of states (LDOS) of the island near the $j$th junction normalized to its normal state value $\nu_N(0)$, index $j=L,R$ stands for the left and the right junctions,  $f_{N}(\varepsilon)=(e^{\varepsilon/T}+1)^{-1}$ is the Fermi distribution function in the normal leads,  and  $f_{S}^{e(o)}(\varepsilon)$ is the distribution function in the S island describing the states with an even (odd) total number of electrons.
The Coulomb blockade effect and the bias voltage $V_{L,R} = \pm V/2$ determine the energy cost
$U_{L,R}=E_C (2 n_g - 1) -e V_{L,R}$ for tunneling.

The increasing magnetic field and vortex entry affect both the LDOS $\nu_{L,R} (\varepsilon)$
and distribution function $f_S^{e,o}(\varepsilon)$ in the above expressions.
To find the distribution function $f_{S}^{e(o)}(\varepsilon)$ we assume that the zero (single) charge state corresponds to an even (odd)
total number of electrons and use the so-called parity projection technique,\cite{Tuominen_Tinkham,Janko_Ambegaokar,Golubev_Zaikin}
\begin{equation}\label{f_S^o,e}
f_{S}^{e(o)}(\varepsilon) =\frac{f_F(\varepsilon)\mp\exp(-2N_{qp})f_B(\varepsilon)}{1\pm\exp(-2N_{qp})} \ ,
\end{equation}
where $f_{F,B}(\varepsilon)=(e^{\varepsilon/T}\pm1)^{-1}$ are the Fermi and Bose distribution functions. The number of quasiparticles can be expressed as
\begin{equation}
N_{qp} = 2 \nu_N(0) \int dV \int_0^\infty \nu(\varepsilon,{\bf r}) f_F(\varepsilon) d\varepsilon \ .
\end{equation}
In the limit $f_{F,B}\ll 1$ one can neglect the difference between these distribution functions and reduce \eqref{f_S^o,e} to the form\cite{2014_Golubev_Maisi_NSN}
$f_S^{e,o}(\varepsilon) = A_{e,o} f_F(\varepsilon)$
with the factor $A_{e}=A_{o}^{-1} = \tanh(N_{qp})$.
In the low temperature limit $T\ll\varepsilon_{\min}$ with $\varepsilon_{\min}$ being the minigap in the quasiparticle spectrum of the island, we obtain $N_{qp} \approx N_{eff} e^{-\varepsilon_{\min}/T}$,
where $N_{eff}$ is a slow function of temperature $T$ (see Appendix~\ref{SM:N_qp} for details).

Within the region of the essential parity effect (when $|A_k-1|\sim 1$) we can rewrite the tunneling rate as follows:
\begin{subequations}\label{Gamma_pm1}
\begin{align}
\Gamma_{j}^k[U>0] &=I_j(U)/e\left[1+\frac{A_k}{e^{U/T}-1}\right] \ , \\
%\frac{I_j(U)/e}{1-e^{-U/T}}\left[1+e^{-U/T}(A_k-1)\right] \ , \\
\Gamma_{j}^k[U<0] &=\frac{I_j(U)/e}{1-e^{-U/T}}A_k \ ,
\end{align}
\end{subequations}
where the ``seed'' IV-characteristic of the tunnel junction in the absence of the Coulomb effects is
\begin{equation}
I_j(U) =\int\limits_{-\infty}^{\infty}\frac{\nu_j(\varepsilon)}{e R_{Tj}} [f_N(\varepsilon-U) -f_N(\varepsilon)]d\varepsilon  \ .
\end{equation}
Note that $I_j(-U)=-I_j(U)$.
 Further calculations should assume a certain model describing the dependence of the IV curves $I_j(U)$
 and the number of quasiparticles $N_{qp}$ on the applied magnetic field.
 For the sake of simplicity we consider the S island to be symmetric (see Fig.~\ref{Fig:setup}) assuming LDOS and tunnel resistances at both junctions to be equal, i.e., $\nu_j=\nu (\varepsilon)$ and $R_{Tj}=R_T$.
 %have the form of a disc of radius $R$. The magnetic field $\bm H$ is perpendicular to the disc plane as shown in Fig.~\ref{Fig:setup}.
 In this case key parameters governing the behavior of the IV curve, i.e., the minigaps $\varepsilon_{j}$ in the quasiparticle
 spectrum at the $j$th junction are also equal $\varepsilon_j=\varepsilon_{edge}$.
%For the chosen geometry and assuming the vortex entering the disc to be
% positioned at the disc center, the LDOSes at both junctions are equal, i.e., $\nu_L=\nu_R=\nu (\varepsilon)$, therefore $\varepsilon_L=\varepsilon_R =\varepsilon_{edge}$.
The most important part of $I_j(U)$ controlling the charge transfer corresponds to small voltages ($U\lesssim\varepsilon_{edge}$) when the IV curve reveals the temperature activated behavior,
 \begin{equation}\label{IV}
I_j(U)\simeq \frac{T}{ eR_{T}}e^{-(\epsilon_{edge}-U)/T}
\int\limits_{0}^{\infty}\nu(\varepsilon_{edge}+Tx)  e^{-x} dx\ .
 \end{equation}
In the large voltage limit ($U\gg\varepsilon_{edge}$) we assume a linear dependence $I_j(U)=U/eR_{T}$.
Note that we neglect here a low voltage contribution to the current arising from the exponential tail of
the residual density of states localized inside the vortex core.

Thus, the basic characteristics of our rate equation are determined by the magnetic field dependence of two energy scales:
(i) the spectral gap $\varepsilon_{edge}$ at the junctions and (ii) the minimal spectral gap $\varepsilon_{\min}$ over the island.
Considering an exemplary geometry shown in Fig.~\ref{Fig:setup} one can see that
the energy scale $\varepsilon_{\min}$  is determined by the maximum of the local superfluid velocity $v_S$
 reached either at the edge of the S disc or in the vortex core.
The gap $\varepsilon_{edge}$ at the junctions is determined by the geometry of the S leads attached to the disc.
Adding these S leads one can control the magnetic field effect on the tunneling DOS and parity phenomenon
independently.
Taking, e.g., the diffusion limit with the coherence length well exceeding the mean free path $\ell$ we find (see Refs.~\onlinecite{1964_Skalsky, 1965_Maki, 2003_DOS_in_H}):
\begin{equation}
\label{edge}
\varepsilon_{edge} = \Delta(H) (1-\gamma_H^{2/3})^{3/2} \ ,
\end{equation}
where $\Delta(H)= \Delta(0) e^{-\pi \gamma_H/4}$,
$\gamma_H = \hbar \langle v_S^2\rangle/[2 D \Delta(H)]$, $D$ is the diffusion coefficient,
${\langle v_S^2\rangle = (\pi DH w/\Phi_0)^2/3}$,
%${\gamma_H = (\Delta (0)/6\Delta(H)) (\pi H w\xi/\Phi_0)^2}$,
 and $w\ll\xi$ is the width of the S lead.
 Estimating now the energy scale $\varepsilon_{\min}$ we can use the same expression \eqref{edge} substituting
 $\gamma_H = \hbar v_S^2/[2 D \Delta(H)]$ with the maximum local superfluid velocity $v_S$.
 Assuming the screening effects to be small, i.e., when the disc radius is smaller than the effective London penetration depth, we get:
$\max v_S = \pi D H R/\Phi_0$.
Considering clean limit $\xi\ll\ell$ we should put $\varepsilon_{\min} = \Delta(H)-\hbar k_F v_S$, where
$\Delta(H)\simeq\Delta(0)[1-\alpha H^2/H_c^2]$.
Here, $H_c\sim\Phi_0/R\xi$ is the field of the first vortex entry and $\alpha$ is
a numerical factor of order unity. Just before the entry of the first vortex  $\varepsilon_{\min}$
tends to zero in the clean limit and remains at finite value in the dirty limit.\cite{Gap_in_H_Fulde,Gap_in_H_Sols,Gap_in_H_Vodolazov}
After the vortex enters
 $\varepsilon_{min}$ equals the minigap $\omega_0$ in the clean limit and turns to zero
in the dirty regime.
%%%%%%%%%%%%%%%%%%%%%%%%%%%%%%%%%%%%%%%%%%%%%%%%%%%%%%%%%%%%%%%%%%%%
\begin{figure}[t]
\includegraphics[width=0.35\textwidth]{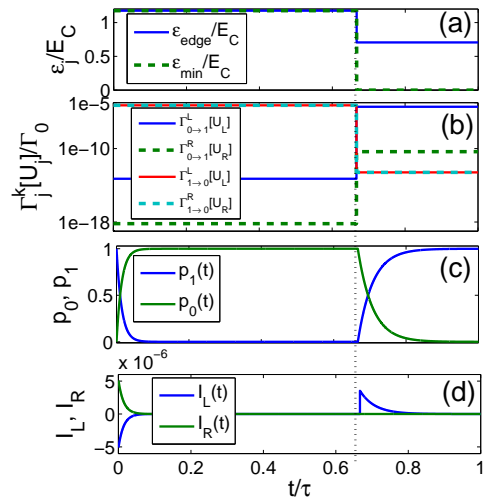}%{Fig3_epsilon_edge.eps}
\caption{(color online) The time-dependence of the energy scales $\varepsilon_{edge}$ and $\varepsilon_{\min}$
(a), the tunneling rates normalized to $\Gamma_0$ (b), the probability distributions (c), and the instantaneous currents through $j$th junction (d).
%The time protocol of the energy scales: the gap at the edge of the island $\varepsilon_{edge}$ (red solid line) and the minimal energy gap over the island $\varepsilon_{\min}$ (blue dashed line).
}
 \label{Fig:protocol}
\end{figure}
%%%%%%%%%%%%%%%%%%%%%%%%%%%%%%%%%%%%%%%%%%%%%%%%%%%%%%%%%%%%%%%%%%%%

In order to model vortex induced pumping we assume the following
time dependence of the spectral gaps with the period $\tau=t_0+t_v$ (see Fig.~\ref{Fig:protocol}) dictated by the piecewise constant magnetic field applied: $\varepsilon_{edge} =\varepsilon_{\min}=\Delta(0)$
for $0<t<t_0$ and $\varepsilon_{edge}(H_m) =\Delta_v$, $\varepsilon_{\min}(H_m)=\varepsilon_0$ in the interval $t_0<t<\tau$.
The characteristic times of the vortex entry/exit are assumed to be negligible comparing to $t_v$ and $t_0$.
Changing the value $\varepsilon_0$ from zero to $\omega_0$ we have the crossover from the dirty to the clean limit (at $H_m>H_c$).
The average current flowing through the $j$-th junction at time instant $t$ can be written as follows
\begin{equation}\label{J_j}
J_{j}(t) = e\left(\Gamma^{e}_j[U_j(t)]p_0(t)-\Gamma^{o}_j[-U_j(t)]p_1(t)\right)
\end{equation}
using the solution of Eq.~\eqref{rate_eq}
\begin{equation}\label{p_n_solution}
p_1(t) = p_1(0)e^{-\int_0^t \Gamma_\Sigma dt'} + \int_0^t \Gamma_{0\to 1}e^{-\int_{t'}^t \Gamma_\Sigma dt''}dt' \ ,
\end{equation}
$p_0(t)=1-p_1(t)$, where $\Gamma_\Sigma = \Gamma_{0\to1}+\Gamma_{1\to 0}$.
In the case of periodic magnetic field protocol $\varepsilon_{edge}(t+\tau)=\varepsilon_{edge}(t)$,  $\varepsilon_{\min}(t+\tau)=\varepsilon_{\min}(t)$ the probability distribution arrives at the periodic steady solution after transient processes, when we can impose the condition $p_1(0)=p_1(\tau)$.
Assuming $U_R=U_L-|eV|>0$, $U_L<\Delta_v$ and $R_L=R_R$ we can simplify the expressions for the tunneling rates \eqref{Gamma_pm1}
\begin{subequations}\label{Gamma_eo}
\begin{align}
\Gamma_{L,R}^{e} &\approx \Gamma_0 e^{-(\varepsilon_{edge}(t)-U_{L,R})/T} \ , \\
\Gamma_{L,R}^{o} &\approx\frac{\Gamma_0 e^{-\varepsilon_{edge}(t)/T}}{\tanh(N_{eff}e^{-\varepsilon_{\min}(t)/T})} \ ,\label{Gamma_10}
\end{align}
\end{subequations}
where we neglect the slow time dependence of the parameter $\Gamma_0 =e^{(\varepsilon_{edge}(t)-U)/T}I_j(U>0,t)/e$ (see Eq.~\eqref{IV}).
Substituting these expressions we find the current averaged over the period $\tau=1/f$,
\begin{multline}\label{J_aver}
\langle J \rangle = \frac{1}{\tau}\int_0^\tau J_j(t) dt \simeq \frac{e f}{2} \left[1 - 2e^{-|eV|/T}+\gamma_0 t_0\right. \\ \left. + \gamma_v t_v- e^{- \Gamma_{\Sigma}^0 t_0}-e^{-\Gamma_{\Sigma}^v t_v}\right] \
\end{multline}
%%%%%%%%%%%%%%%%%%%%%%%%%%%%%%%%%%%%%%%%%%%%%%%%%%%%%%%%%%%%%%%%%%%%
\begin{figure}[t]
%\centering{
\includegraphics[width=0.23\textwidth]{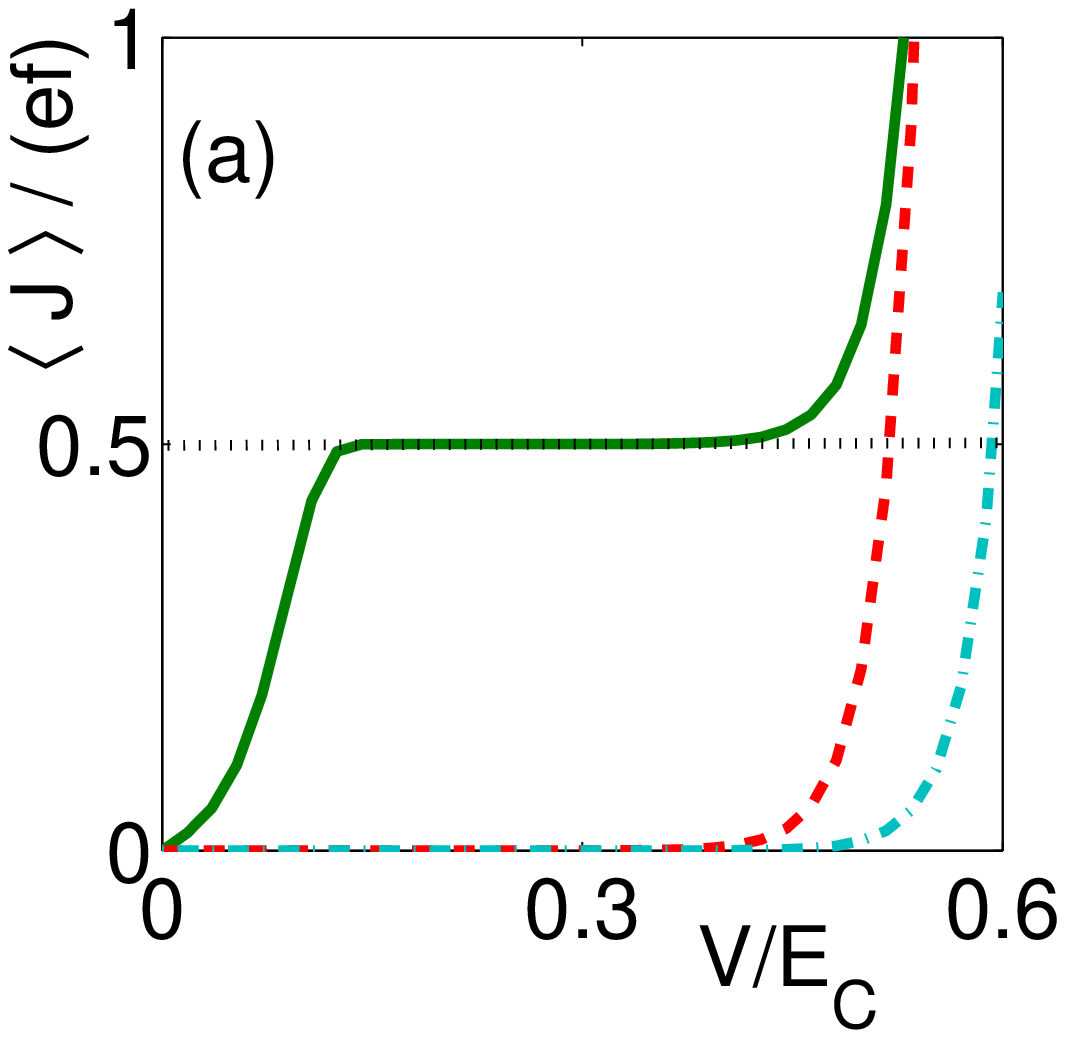}
\includegraphics[width=0.23\textwidth]{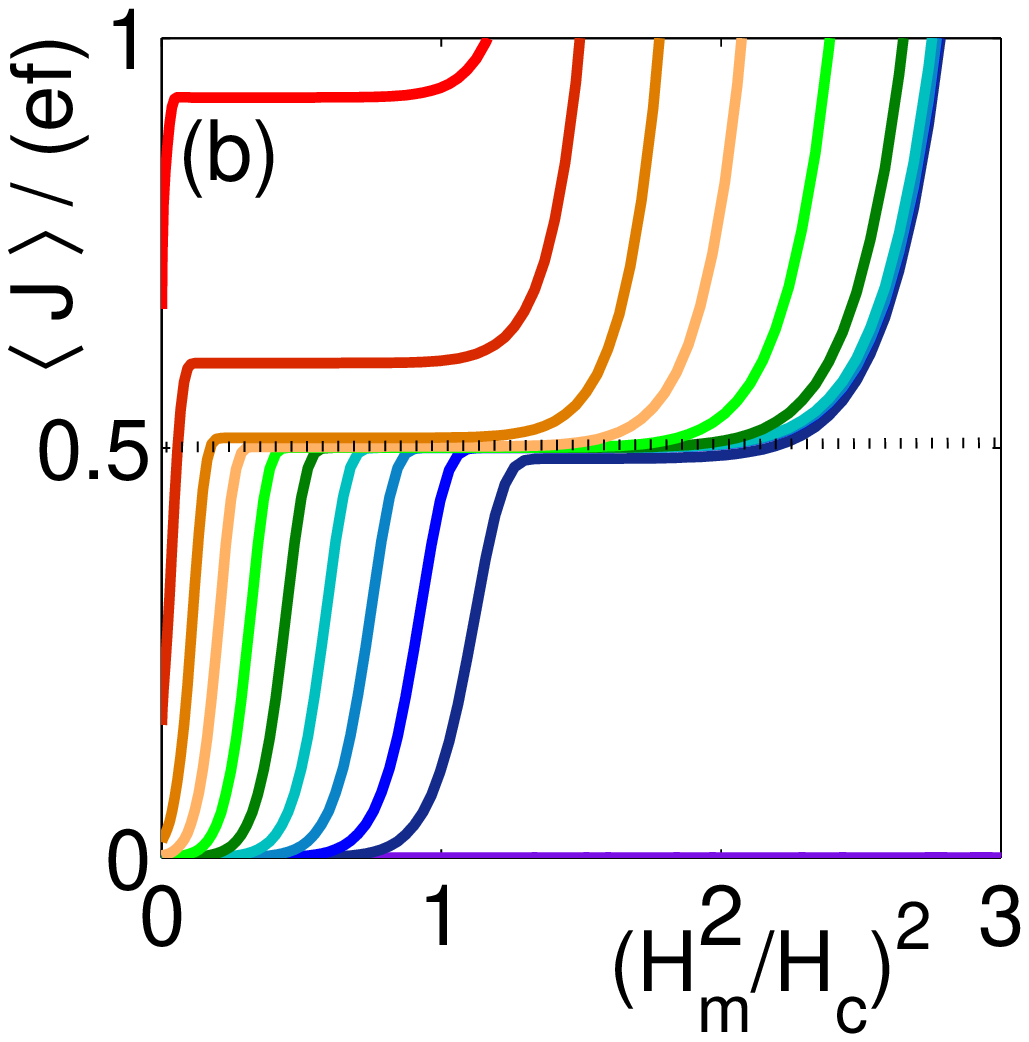}
%}
\caption{(color online) (a) The current $\langle J\rangle$ averaged over the period vs $V$ for $H_m=H_c$ (green solid line) and IV curves at zero $H=0$ (red dashed) and maximal $H=H_m$ (blue dash-dotted) magnetic fields;
(b) The current $\langle J\rangle$ vs squared amplitude of ac magnetic field $H_m$ normalized to the field of the first vortex entry $H_c$ at $V = 0 .. 0.5 E_C$ from left to right.
}
 \label{Fig:J_aver}
\end{figure}
%%%%%%%%%%%%%%%%%%%%%%%%%%%%%%%%%%%%%%%%%%%%%%%%%%%%%%%%%%%%%%%%%%%%
shown in Fig.~\ref{Fig:J_aver}(a) and (b) vs bias voltage $V$ and the magnetic field amplitude $H_m$ in the time interval $t_0<t<\tau$.
Here the total tunneling rates $\Gamma_\Sigma^{v}$ and $\Gamma_\Sigma^{0}$  in the vortex and Meissner states are determined mostly by the maximal rates $\Gamma_\Sigma^0\approx 2\Gamma_0/N_{eff}$, $\Gamma_\Sigma^v \approx \Gamma_0 e^{-(\Delta_v-U_L)/T}$,
while $\gamma_0 = \Gamma_0 e^{-(\Delta-U_L)/T}$, $\gamma_v = \frac{2\Gamma_0 e^{-\Delta_v/T}}{\tanh(N_{eff}e^{-\varepsilon_{0}/T})}$ are the small leakage rates.
The first two terms in the above expression can be obtained from the following reasoning.
When the vortex enters the S island
the total charge transmitted through both junctions should be equal to the electron charge.
Due to the large ratio $\Gamma_{L}^{e}/\Gamma_{R}^{e} = e^{|eV|/T}\gg 1$ of the tunneling rates
the most of this charge transfer $e(1-\exp(-|eV|/T))$ occurs through the left junction,
while only the exponentially small part of it $\sim e \exp(-|eV|/T)$ is transmitted through the right junction.
The vortex exit should be accompanied by the discharge of the island which occurs with equal rates through the both junctions \eqref{Gamma_10}.
As a result, half of electron charge exits the island through each junction.  Summing up the total charge
transmitted through the system per cycle we find $Q=e(1/2 - e^{-|eV|/T})$.
The above symmetry of the discharging processes results in a rather strong shot noise in the system:
the fluctuating transmitted charge equals to $e/2$ and the resulting current noise is given by the expression
$\sqrt{\langle\delta J^2\rangle} = ef/2$.
The last two terms in \eqref{J_aver} appear if the time intervals of two stages $t_v$ and $t_0$ (with and without vortex, respectively) become comparable
or shorter than the characteristic charging times $1/\Gamma_\Sigma^{v}$ and $1/\Gamma_\Sigma^{0}$.
Therefore the maximum operation frequency $f=1/\tau$ is limited by a single quasiparticle tunnel rate $\Gamma_0/N_{eff}$.
Besides the effect of the frequency the average current $\langle J\rangle$ also deviates from $ef/2$ at small bias voltages and/or small magnetic field amplitudes, due to the dependence of total rates $\Gamma_\Sigma^k$ on these parameters (see Fig.~\ref{Fig:J_aver}).
The terms proportional to $\gamma_0$ and $\gamma_v$ originate from the leakages and lead to currents exceeding $ef/2$ shown at Fig.~\ref{Fig:J_aver} for larger $V$ and/or $H_m$ values.
Nevertheless satisfying  the conditions
\begin{subequations}\label{t_m_conditions}
\begin{align}
N_{eff}/\Gamma_0 &\ll t_0 \ll \gamma_0^{-1} \ ,\\
e^{(\Delta_v-U_L)/T}/\Gamma_0 &\ll t_v \ll \gamma_v^{-1} \ ,
\end{align}
\end{subequations}
one can obtain the plateau of the average current at $e f / 2$ (see Appendix~\ref{SM:parameter_range} for details).
These conditions can be met provided we set $eV\gg T$, $\min(T,\varepsilon_0-T\ln N_{eff})\ll U_L\ll \Delta-T\ln N_{eff}$.
%, and $U_L\gg T\ln[\coth(N_{eff} e^{-\varepsilon_0/T})]$.
The breakdown of the above condition at the lower bound of $U_L$ can signal the presence of the minigap in the vortex core in the clean limit $\varepsilon_0 = \omega_0$.
Note that for non-symmetric case, the resulting average current will deviate from $ef/2$.
However, a finite averaged current ranging between $0$ and $ef$ can be obtained in this case as well.

To sum up, we have studied the interplay between vorticity and electric charge which can manifest in conditions of Coulomb
blockade through the vortex induced suppression of the parity effect in mesoscopic samples.
Vortex entry and exit from the sample is shown to be accompanied by synchronized entry and exit of a single electron charge.
Applying the bias voltage and oscillating magnetic field one can observe a vortex governed turnstile phenomenon:
the switching between the Meissner and vortex states periodically opens the device for single charge transfer. Thus, we have demonstrated
that the SET devices provide a unique tool for manipulating the collective dynamics
of charge and vorticity in mesoscopic superconducting samples.

\acknowledgements
We are grateful to D.~Yu.~Vodolazov for useful comments.
This work has been supported in part by Academy of Finland though its LTQ CoE grant
(project no. 250280), the European Union Seventh Framework Programme INFERNOS (FP7/2007-2013) under Grant Agreement No. 308850,  by the Russian Foundation for Basic Research,
the Russian president foundation (SP-1491.2012.5), and the grant of the Russian Ministry of Science and Education No.
02.B.49.21.0003.

\appendix

\section{Derivation of expression for $N_{qp}$ and of Eqs.~\eqref{Gamma_pm1}}\label{SM:N_qp}
In the low temperature limit $T\ll\varepsilon_{\min}$ with $\varepsilon_{\min}$ being the minigap in the quasiparticle spectrum of the island Eq.~\eqref{N_qp} from main text can be written as follows
\begin{equation}
N_{qp} \approx N_{eff} e^{-\varepsilon_{\min}/T} \ ,
\end{equation}
where
\begin{equation}
N_{eff} = 2 \nu_N(0)T \int dV \int_0^\infty \nu(\varepsilon_{\min}+T\cdot x,{\bf r}) e^{-x} dx
\end{equation}
is a slow function of temperature $T$. Indeed,
\begin{multline}
\frac{\partial N_{eff}}{\partial T} = 2 \nu_N(0)\int dV \int_0^\infty x \nu(\varepsilon_{\min}+T\cdot x,{\bf r}) e^{-x} dx\simeq\\
\frac{N_{eff}}{T}\ll \frac{N_{eff}}{T}\frac{\varepsilon_{\min}}{T} \ .
\end{multline}

To derive Eqs.~\eqref{Gamma_pm1} from the main text it is convenient to rewrite Eq.~\eqref{Gamma_pm} for the tunneling rates in the form:
\begin{multline}
\Gamma_{j}^k[U] =\frac{1}{e^2 R_{Tj}}\int\limits_{-\infty}^{\infty}\nu_j(\varepsilon) f_N(\varepsilon-U) [1-f_N(\varepsilon)]d\varepsilon \\
+\frac{(A_k-1)}{e^2 R_{Tj}}\int\limits_{0}^{\infty}\nu_j(\varepsilon)  f_N(\varepsilon) [1-f_N(\varepsilon-U)-f_N(\varepsilon+U)] \ ,
\end{multline}
where we explicitly separate the parity effect contribution.

These expressions read
\begin{multline}
\Gamma_{j}^k[U] =\frac{I_j(U)/e}{1-e^{-U/T}} \\
+(A_k-1)\left[\frac{I_j(|U|)/e}{e^{|U|/T}-1}-\Gamma_{qp} e^{-|U|/T}\right] \ ,
\end{multline}
where $I_j(U)$ is the ``seed'' IV-characteristic of the tunnel junction in the absence of the Coulomb effects given by the Eq.~\eqref{seed_IV} of the main text and
%\begin{equation}
%I_j(U) =\int\limits_{-\infty}^{\infty}\frac{\nu_j(\varepsilon)}{e R_{Tj}} [f_N(\varepsilon-U) -f_N(\varepsilon)]d\varepsilon  \ ,
%\end{equation}
\begin{equation}
\Gamma_{qp}=\frac{T}{e^2 R_{Tj}}e^{-\varepsilon_j/T}\int\limits_{0}^{\infty}\nu_j(\varepsilon_j+Tx)  e^{-x} dx \ .
\end{equation}

Within the region of the essential parity effect (when $|A_k-1|\sim 1$) we can obviously neglect the term proportional to $\Gamma_{qp}$ and obtain the Eqs.~\eqref{Gamma_pm1} from the main text:
\begin{subequations}
\begin{align}
\Gamma_{j}^k[U>0] &=I_j(U)/e\left[1+\frac{A_k}{e^{U/T}-1}\right] \ , \\
%\frac{I_j(U)/e}{1-e^{-U/T}}\left[1+e^{-U/T}(A_k-1)\right] \ , \\
\Gamma_{j}^k[U<0] &=\frac{I_j(U)/e}{1-e^{-U/T}}A_k \ .
\end{align}
\end{subequations}

\section{Derivation of Eq.~\eqref{J_aver}}\label{SM:J_aver}
Using the magnetic field protocol considered in the main text $\varepsilon_{edge} =\varepsilon_{min}=\Delta(0)$
for $0<t<t_0$ and $\varepsilon_{edge}(H_m) =\Delta_v$, $\varepsilon_{min}(H_m)=\varepsilon_0$ in the interval $t_0<t<\tau$, and the periodicity condition
\begin{equation}
p_1(0)=p_1(\tau) = \frac{\int_0^\tau \Gamma_{0\to 1}e^{-\int_{t'}^\tau \Gamma_\Sigma dt''}dt'}{1-e^{-\int_0^\tau \Gamma_\Sigma dt'}} \ .
\end{equation}
one can rewrite the solution~\eqref{p_n_solution} in the main text as follows
\begin{gather}
p_1(t<t_0) = p_{1,eq}^0+\left(p_{1,eq}^v-p_{1,eq}^0\right)\frac{e^{-\Gamma_{\Sigma}^0 t}\left(1-e^{-\Gamma_{\Sigma}^v t_v}\right)}{1-e^{-\Gamma_{\Sigma}^0 t_0-\Gamma_{\Sigma}^v t_v}} \ , \\
p_1(t>t_0) = p_{1,eq}^v+\left(p_{1,eq}^0-p_{1,eq}^v\right)\frac{e^{-\Gamma_{\Sigma}^v(t-t_0)}\left(1-e^{-\Gamma_{\Sigma}^0 t_0}\right)}{1-e^{-\Gamma_{\Sigma}^0 t_0-\Gamma_{\Sigma}^v t_v}} \ .
\end{gather}
Here the superscript $0$ ($v$) corresponds to the time interval $0<t<t_0$ ($t_0<t<\tau$, $\tau=t_0+t_v$) and $p_{1,eq}^k = \Gamma_{0\to 1}^k/\Gamma_{\Sigma}^k$ is the adiabatic solution in the corresponding time interval.

Substituting this solution to Eqs.~(\ref{J_j}, \ref{J_aver}) from the main text one can obtain
\begin{multline}\label{SM:I_aver}
\langle J \rangle\tau = J_{eq}^0 t_0 + J_{eq}^v t_v + e\left(\frac{\Gamma_R^v}{\Gamma_\Sigma^v}-\frac{\Gamma_R^0}{\Gamma_\Sigma^0}\right)\times\\\left(p_{1,eq}^v-p_{1,eq}^0\right)\frac{\left(1-e^{-\Gamma_{\Sigma}^0 t_0}\right)\left(1-e^{-\Gamma_{\Sigma}^v t_v}\right)}{1-e^{-\Gamma_{\Sigma}^0 t_0-\Gamma_{\Sigma}^v t_v}}
\end{multline}
where $\Gamma_R^k = \Gamma_{0\to 1}^{R,k}+\Gamma_{1\to 0}^{R,k}$ and $J_{eq}^k = e (\Gamma_{0\to 1}^{R,k}\Gamma_{1\to 0}^{L,k}-\Gamma_{0\to 1}^{L,k}\Gamma_{1\to 0}^{R,k})/\Gamma_\Sigma^k$ is the current flowing through the island when the probabilities follow the adiabatic solution $p_{eq}^k$.

Using the expressions for all tunneling rates, assuming the conditions of Eqs.~\eqref{t_m_conditions} from the main text to be valid and keeping only the first order corrections in small parameters $e^{-|eV|/T}$, $\gamma_0 t_0$, $\gamma_v t_v$, $e^{-\Gamma_\Sigma^0 t_0}$, and $e^{-\Gamma_\Sigma^v t_v}$ we obtain
\begin{gather}
p_{eq}^0\approx \frac{\gamma_0}{\Gamma_\Sigma^0}, \quad p_{eq}^v\approx 1-\frac{\gamma_v}{\Gamma_\Sigma^v}, \quad J_{eq}^k\approx e\frac{\gamma_k}{2},\\
\frac{\Gamma_R^v}{\Gamma_\Sigma^v}\approx 1-\frac{\gamma_v}{2\Gamma_\Sigma^v}-e^{-|eV|/T}, \quad \frac{\Gamma_R^0}{\Gamma_\Sigma^0}\approx \frac{1}{2}+\frac{\gamma_0}{2\Gamma_\Sigma^0}.
\end{gather}
Substituting these expressions into \eqref{SM:I_aver} we get Eq.~\eqref{J_aver} from the main text.

\section{Range of parameters for current plateau observation}\label{SM:parameter_range}
In this section we consider ranges of the parameters where the plateau of the current $J_j(t)$ averaged over the period $\tau$ is close to $ef/2$, i.e., $\langle J\rangle =ef(1\pm \epsilon)/2$ with a certain small $\epsilon\ll 1$.
Using the result \eqref{J_aver} from the main text one can roughly rewrite this condition as follows
\begin{gather}
e^{-|eV|/T}, \gamma_0 t_0, \gamma_v t_v, e^{-\Gamma_\Sigma^0 t_0}, e^{-\Gamma_\Sigma^v t_v}\lesssim \epsilon \ .
\end{gather}

Further we focus on IV characteristic for $\langle J\rangle(V)$ assuming that all other parameters ($T$, $\Delta$, $\Delta_v$, $\varepsilon_0$, $E_{ch}=E_C(2n_g-1)$, $N_{eff}$, $t_0$, and $t_v$) are chosen to be optimal for rather small bias voltages.
%Let's consider first the case $N_{eff}e^{-\varepsilon_0/T}\gg 1$.

Considering all the corrections to the current plateau one can separate them into two groups:
(i) voltage-independent corrections
\begin{gather}\label{SM:conds_v_indep}
2\Gamma_0 t_0/N_{eff} > \ln(1/\epsilon), \quad \frac{2\Gamma_0 e^{-\Delta_v/T} t_v}{\tanh(N_{eff}e^{-\varepsilon_{0}/T})} <\epsilon,
\end{gather}
and the necessary condition at $V=0$
\begin{gather}
\Gamma_0 e^{-(\Delta-E_{ch})/T} t_0 <\epsilon, %\\ \Gamma_0 e^{-(\Delta_v-E_{ch})/T} t_v>\ln(1/\epsilon), \quad
\end{gather}
(ii) voltage-dependent corrections
\begin{gather}
e^{-|eV|/T}<\epsilon, \quad \Gamma_0 e^{-(\Delta-E_{ch}-|eV|)/T} t_0 <\epsilon, \nonumber\\
\label{SM:conds}
\Gamma_0 e^{-(\Delta_v-E_{ch}-|eV|)/T} t_v>\ln(1/\epsilon), \\
\Gamma_0 e^{-(\Delta_v+E_{ch}-|eV|)/T} t_v<\epsilon,\nonumber
\end{gather}
which can be rewritten as the conditions on the bias voltage
\begin{subequations}\label{SM:V_ranges}
\begin{align}
|eV|&>T \ln(1/\epsilon) \ , \\
|eV|&>\Delta_v-E_{ch}-T\ln[\Gamma_0 t_v/\ln(1/\epsilon)] \ , \\
|eV|&<\Delta-E_{ch}-T\ln[\Gamma_0 t_0/\epsilon] \ , \\
|eV|&<\Delta_v + E_{ch}-T\ln[\Gamma_0 t_v/\ln(1/\epsilon)] \ .
\end{align}
\end{subequations}
The last term in \eqref{SM:conds}  originates from the condition $\gamma_v t_v\lesssim \epsilon$ for the case $U_R<-T\ln[1/\tanh(N_{eff}e^{-\varepsilon_{0}/T})]<0$.

One can see that the increase in the minigap $\varepsilon_0$ modeling crossover between vortex minigaps in dirty and clean limits breaks first the last voltage-independent condition in Eq.~\eqref{SM:conds_v_indep}.
As a result, the plateau of the averaged current will be shifted to $\gamma_v t_v$ as a whole without change in the range of the bias voltage \eqref{SM:V_ranges}.

\end{document}